\documentclass[12pt]{iopart}

\usepackage{graphicx}% Include figure files
\usepackage{bm}% bold math
\usepackage{dcolumn}% Align table columns on decimal point
\usepackage{float}
\usepackage{citesort}
\usepackage{contigrefs}

\newcommand{\ket}[1]{\ensuremath{\left|#1\right\rangle}}

\begin{document}

\title{Hyperfine and quadrupole interactions for Dy isotopes in DyPc$_2$ molecules}

\author{Aleksander L. Wysocki and Kyungwha Park}
\address{Department of Physics, Virginia Tech, Blacksburg, Virginia 24061, United States}
\ead{alexwysocki2@gmail.com; kyungwha@vt.edu}

\begin{abstract}
Nuclear spin levels play an important role in understanding magnetization dynamics and implementation and control of quantum bits in lanthanide-based single-molecule magnets. We investigate the hyperfine and nuclear quadrupole interactions for $^{161}$Dy and $^{163}$Dy nucleus in anionic DyPc$_2$ (Pc=phthalocyanine) single-molecule magnets, using multiconfigurational {\it ab-initio} methods (beyond density-functional theory) including spin-orbit interaction. The two isotopes of Dy are chosen because the others have zero nuclear spin. Both isotopes have the nuclear spin $I=5/2$, although the magnitude and sign of the nuclear magnetic moment differ from each other. The large energy gap between the electronic ground and first-excited Kramers doublets, allows us to map the microscopic hyperfine and quadrupole interaction Hamiltonian onto an effective Hamiltonian with an electronic pseudo-spin $S_{\rm eff}=1/2$ that corresponds to the ground Kramers doublet. Our {\it ab-initio} calculations show that the coupling between the nuclear spin and electronic orbital angular momentum contributes the most to the hyperfine interaction and that both the hyperfine and nuclear quadrupole interactions for $^{161}$Dy and $^{163}$Dy nucleus are much smaller than those for $^{159}$Tb nucleus in TbPc$_2$ single-molecule magnets. The calculated separations of the electronic-nuclear levels are comparable to experimental data reported for $^{163}$DyPc$_2$. We demonstrate that hyperfine interaction for Dy Kramers ion leads to tunnel splitting (or quantum tunneling of magnetization) at zero field. This effect does not occur for TbPc$_2$ single-molecule magnets. The magnetic field values of the avoided level crossings for $^{161}$DyPc$_2$ and $^{163}$DyPc$_2$ are found to be noticeably different, which can be observed from experiment.
\end{abstract}

\vspace{2pc}
\noindent{\it Keywords}: lanthanide single-molecule magnets, hyperfine coupling, nuclear quadrupole interaction, spin-orbit interaction, ab-initio calculations.


%\maketitle


\section{Introduction}
Lanthanide-based single-molecule magnets (SMMs) \cite{Kasuga1980,Ishikawa2002,Ishikawa2003,Sessoli2009,Reinhart2011,Baldovi2012,Woodruff2013,Liddle2015,Powell2015,Wang2016,Demir2017,Guo2018,Goodwin2017}
have shown a promising possibility for quantum information science applications \cite{Thiele2014,Shiddiq2016,Pedersen2016,Godfrin2017,Godfrin2018,Hussain2018,Arino2019,Wernsdorfer2019,Najafi2019}. Among various routes to realize quantum bits (qubits) or quantum $d$-levels (qudits), utilization of molecular electronic or nuclear spin levels is unique because of large internal and external degrees of freedom for tailoring their properties by varying chemical environmental factors. Recently, Rabi oscillations of nuclear spin levels \cite{Thiele2014} and their applications to quantum algorithms \cite{Godfrin2017} have been experimentally realized in terbium (Tb) based
double-decker SMMs such as TbPc$_2$ (Pc=phthalocyanine) \cite{Ishikawa2003}. Furthermore, the possibility of strong coupling between the nuclear
spin qubits of TbPc$_2$ SMMs via a superconducting resonator was theoretically proposed \cite{Najafi2019}.

\begin{figure}[t!]
\centering
\includegraphics[width=0.4\linewidth]{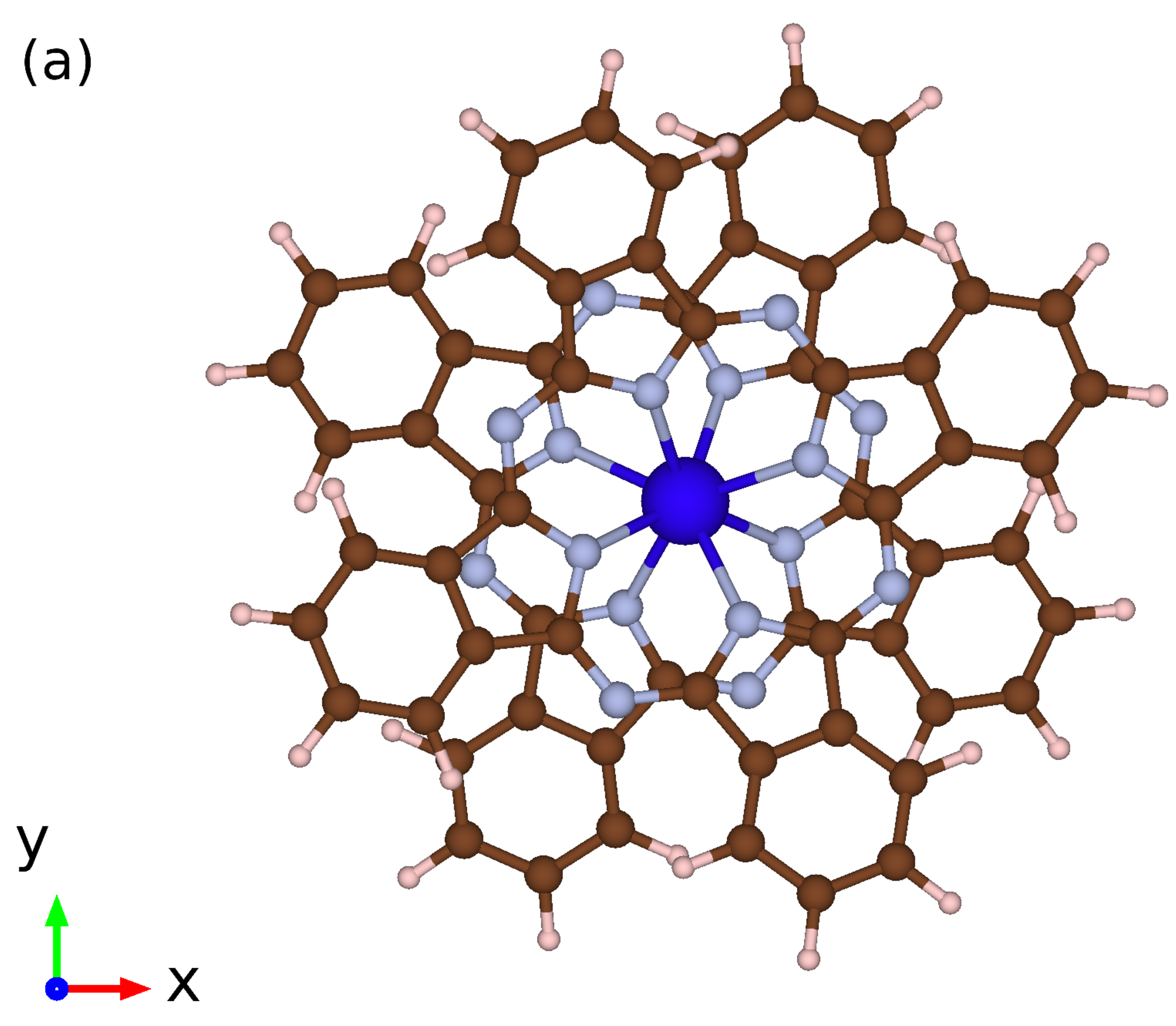}
\hspace{0.5truecm}
\includegraphics[width=0.4\linewidth]{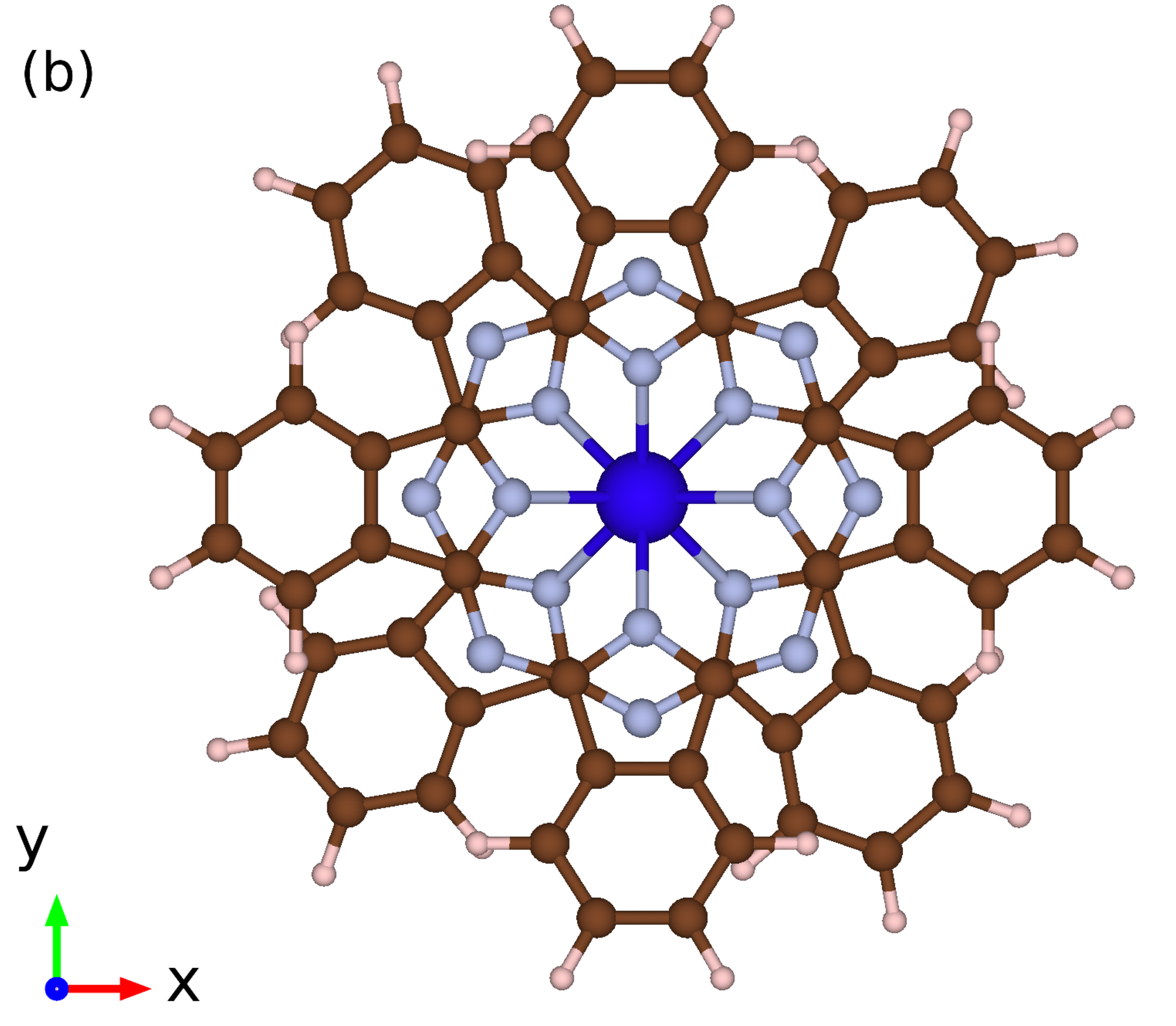}
\vspace{0.8truecm}
\includegraphics[width=0.4\linewidth]{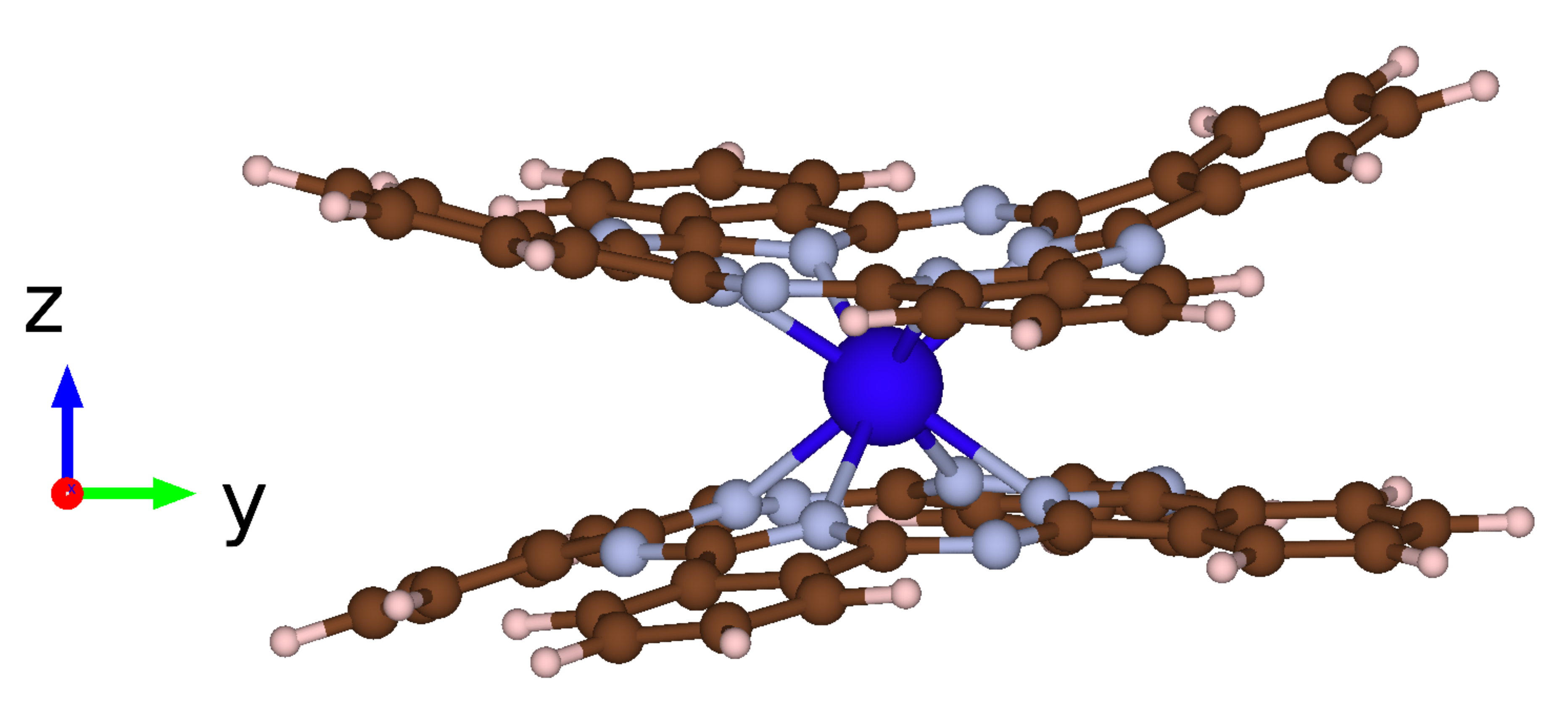}
\hspace{0.5truecm}
\includegraphics[width=0.4\linewidth]{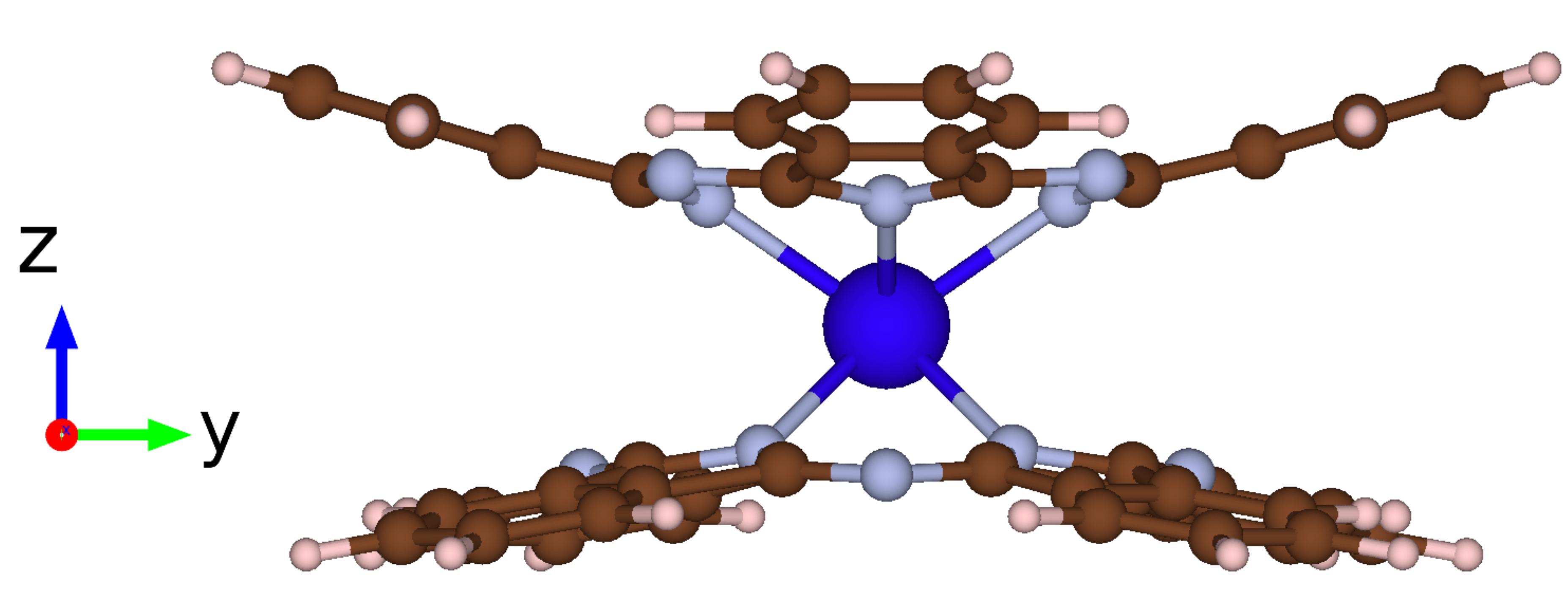}
\caption{Top and side views of the experimental atomic structure of anionic DyPc$_2$ molecules from Ref.~\cite{Marx2014} (a) and
Ref.~\cite{Pineda2017} (b). The molecule in (a) does not have any symmetry, whereas the molecule in (b) has exact $C_4$ symmetry.
Blue, gray, maroon, and pale pink spheres represent Dy, N, C, and H atoms, respectively. The coordinate system corresponds to
magnetic axes obtained by diagonalization of the $\mathbf{g}$-matrix calculated for the electronic ground Kramers doublet for
each molecule. The magnetic easy axis coincides with the $z$ axis.}
\label{Geometry}
\end{figure}

In a TbPc$_2$ SMM, a Tb$^{3+}$ (4$f^8$) ion with the spin angular momentum $S=3$ and the orbital angular momentum $L=3$ is sandwiched between two Pc ligand planes \cite{Ishikawa2003}. A singly charged TbPc$_2$ SMM has the total angular momentum $J=6$ in the ground state with large magnetic anisotropy. $^{159}$Tb isotope has natural abundance of 100\% \cite{Rosman1998} with the nuclear spin $I=3/2$. Multiconfigurational {\it ab-initio} studies showed that the energy gap between the electronic ground and first-excited quasi-doublet is about 300~cm$^{-1}$ \cite{Ungur2017,Pederson2019}, and that the $^{159}$Tb nuclear spin is strongly coupled to the electronic orbital and spin degrees of freedom with hyperfine coupling constant $A_{zz}\sim$500~MHz for the electronic ground quasi-doublet $J_z=\pm 6$ \cite{Wysocki2019}, where the $z$ axis coincides with the magnetic easy axis. This result agrees with the experimental data \cite{Ishikawa2005}.

\begin{figure}[t!]
\centering
\includegraphics[width=0.5\linewidth]{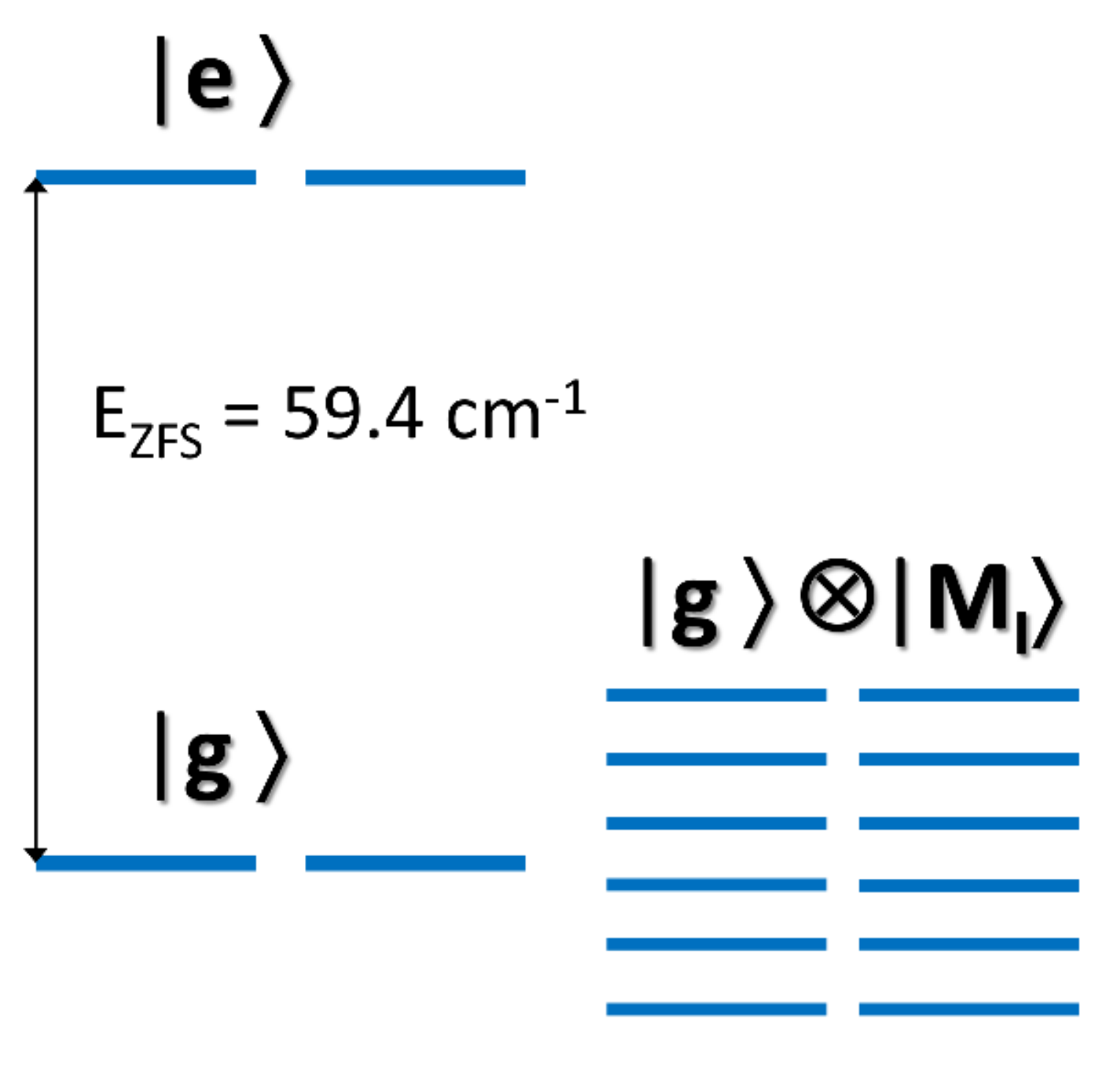}
\caption{Schematic diagram of the low-energy electronic-nuclear energy spectrum for anionic DyPc$_2$ SMMs \cite{Marx2014,Pineda2017}. The electronic ground doublet ($|g \rangle$) is separated from the electronic first-excited doublet ($|e \rangle$) by 59.4 (51.1)~cm$^{-1}$ for Fig.~\ref{Geometry}(a) [(b)]. The doublet $|g \rangle$ consists of primarily $|M_J=\pm13/2\rangle$, whereas the doublet $|e \rangle$ comprises mainly $|M_J=\pm11/2\rangle$. With the hyperfine interactions, each electronic level splits into six (quasi)-doublets $|g \rangle \otimes |M_I \rangle$, considering the Dy nuclear spin $I=5/2$. The separations of the nuclear levels are much less than 0.1~cm$^{-1}$.}
\label{Schematic}
\end{figure}

As an alternative to TbPc$_2$ SMMs, other LnPc$_2$ complexes (Ln=Nd, Dy, Ho, Er, Tm, and Yb) \cite{Kasuga1980,Ishikawa2003,Woodruff2013,Candini2016} can be considered for quantum information science applications. Among them, DyPc$_2$ SMMs (Fig.~\ref{Geometry}) have some advantages over TbPc$_2$ SMMs. Dy element has two different isotopes with non-zero nuclear spin. $^{161}$Dy and $^{163}$Dy have natural abundance of 18.9 and 24.9\%, respectively \cite{Rosman1998}. Both of them have the nuclear spin $I=5/2$. The larger nuclear spin suggests more nuclear spin levels that can be used and controlled for quantum information applications. Furthermore, Dy ions are more susceptible to ligand fields such that the effective magnetic anisotropy barrier can be enhanced by more than two orders of magnitude by varying the surrounding ligands. For example, recently, Dy-based SMMs exhibited magnetic hysteresis above liquid nitrogen temperature \cite{Guo2018} and effective magnetic anisotropy barrier over 1000~cm$^{-1}$ \cite{Goodwin2017}.

In a charged DyPc$_2$ SMM, the Dy$^{3+}$ (4$f^9$) ion has $S=5/2$ and $L=5$, giving rise to $J=15/2$ according to Hund's rules, which is confirmed by {\it ab-initio} calculations \cite{Marx2014}. Thus, the Kramers theorem is applied to the DyPc$_2$ SMM, which is not the case for charged TbPc$_2$ SMMs. The crystal field of the Pc ligands splits the ground $J=15/2$ multiplet into eight Kramers doublets.  Multiconfigurational calculations found that the energy gap between the electronic ground and first-excited doublet ($E_{\rm ZFS}$) for the DyPc$_2$ SMM is about 60~cm$^{-1}$ \cite{Marx2014} (Fig.~\ref{Schematic}), which is about a fifth of the corresponding value for the TbPc$_2$ SMM \cite{Ungur2017,Pederson2019}. Considering the hyperfine and nuclear quadrupole interactions for $^{161}$Dy and $^{163}$Dy nuclei, each electronic level is split into six (quasi)-doublets. Different isotopes have different magnitude and sign of the nuclear magnetic moment, which makes interpretation of experimental data \cite{Ishikawa2005} difficult. So far, there are no {\it ab-initio} studies of the hyperfine interactions of
$^{161}$DyPc$_2$ and $^{163}$DyPc$_2$ SMMs.

In this work, we investigate the hyperfine and nuclear quadrupole interactions of anionic $^{161}$DyPc$_2$ and $^{163}$DyPc$_2$ SMMs, using multiconfigurational {\it ab-initio} methods including spin-orbit interaction (SOI) in comparison to those for $^{159}$TbPc$_2$ SMMs. The hyperfine and quadrupole interactions are considered in the non-relativistic limit. Considering both asymmetric and $C_4$ symmetric experimental geometries~\cite{Marx2014,Pineda2017} (Fig.~\ref{Geometry}), we first identify electronic Kramers doublet structures of the molecules. Then we extract the hyperfine and quadrupole parameters for $^{161}$Dy and $^{163}$Dy nuclei projected onto the electronic ground doublet. Next, the electronic-nuclear levels for both Dy isotopes are obtained and compared with experimental data for $^{163}$DyPc$_2$ \cite{Pineda2017}. There are no reported experimental data for $^{161}$DyPc$_2$. Furthermore, we discuss important consequences of the electronic Kramers doublet coupled to the half-integer nuclear spin on zero-field tunneling splitting and Zeeman diagram.

\section{Methodology and Computational Details}

We use SI units and a magnetic coordinate system where the ${\mathbf g}$ matrix for the electronic ground doublet is diagonal. The methodology used in this work was presented in detail in Refs.~\cite{Wysocki2019,Bolvin2014,Sharkas2015}. Thus, here we briefly explain only the key points and specifics for the $^{161}$DyPc$_2$ and $^{163}$DyPc$_2$ SMMs.

\subsection{Methodology}

The hyperfine interactions consist of three components \cite{AbragamBook}: (i) the coupling between the nuclear spin and the electronic orbital angular momentum; (ii) the dipolar interaction between the nuclear spin and the electronic spin; (iii) the contact interaction between the nuclear spin and the electron spin density at the nucleus position. The first, second, and third components are referred to as the paramagnetic spin-orbital (PSO) contribution, the spin-dipole (SD) interaction, and the Fermi contact (FC) term, respectively. The microscopic hyperfine Hamiltonian $\hat{H}_{\text{MHf}}$ contains all three components. The hyperfine interactions are treated in the non-relativistic limit. When the electronic ground doublet is well separated from the electronic first-excited doublet (Fig.~\ref{Schematic}), the effective hyperfine Hamiltonian $\hat{H}_A$ for an electronic pseudo-spin $S=1/2$ can be described as:
\begin{eqnarray}
\hat{H}_{A}&=&\hat{\mathbf{I}}\cdot\mathbf{A}\cdot\hat{\mathbf{S}},
\label{eq:HA} \\
           &=&\frac{A_{zz}}{2}\hat{I}_z\hat{S}_z + \frac{A_0}{2}\hat{I}_+\hat{S}_- +A_1\left(\hat{I}_-\hat{S}_z+\hat{I}_z\hat{S}_-\right) +\frac{A_2}{2}\hat{I}_-\hat{S}_- + \text{h.c.},
\label{eq:HA2}
\end{eqnarray}
where ${\mathbf A}$ is the magnetic hyperfine matrix, $A_0=\frac{1}{2}\left(A_{xx}+A_{yy}\right)$, $A_1=\frac{1}{2}\left(A_{xz}+iA_{yz}\right)$, and $A_2=\frac{1}{2}\left(A_{xx}-A_{yy}\right)+iA_{xy}$. For an electronic pseudo-spin $S=1/2$, one can relate $\hat{H}_{\text{MHf}}$ to $\hat{H}_A$ by using \cite{Bolvin2014,Sharkas2015}
\begin{equation}
(\mathbf{A}\mathbf{A}^T)_{\alpha\beta}=2\sum_{ij}\langle i|\hat{h}_{\text{MHf}}^\alpha|j\rangle\langle j|\hat{h}_{\text{MHf}}^\beta|i\rangle,
\label{eq:Aformula}
\end{equation}
where $\alpha, \beta=x,y,z$ and $\hat{h}_{\text{MHf}}^\alpha\equiv\partial\hat{H}_{\text{MHf}}/\partial\hat{I}_\alpha$. The summation runs over the {\it ab initio} states of the electronic ground doublet ($i=1,2$). For the $^{161}$Dy nucleus, the nuclear magnetic moment ${\mathbf m}_N$ (=$g_N \mu_N {\mathbf I}$) is antiparallel to ${\mathbf I}$ and the nuclear $g$-factor, $g_N$, is $-$0.19224 \cite{Mills1988}, where $\mu_N$ is the nuclear magneton. For the $^{163}$Dy nucleus, ${\mathbf m}_N$ is parallel to ${\mathbf I}$ and the nuclear $g$-factor is 0.26904 \cite{Mills1988}. The sign of ${\mathbf A}$ cannot be determined from this approach, and so it is chosen from experiment. For example, the experimental data for $^{163}$DyPc$_2$ \cite{Pineda2017} indicates a positive sign for $A_{zz}$, and so we choose that $A_{zz} > 0$ for $^{163}$Dy isotope, while $A_{zz} < 0$ for $^{161}$Dy isotope.

The nuclear quadrupole interaction Hamiltonian is described by
\begin{eqnarray}
\hat{H}_{\text Q}&=&\hat{\mathbf{I}}\cdot\mathbf{P}\cdot\hat{\mathbf{I}}, \label{eq:HQ} \\
&=&\frac{3}{4}P_{zz}\left[\hat{I}_z^2-\frac{I(I+1)}{3}\right]+P_1\left(\hat{I}_z\hat{I}_-+\hat{I}_-\hat{I}_z\right)+\frac{P_2}{2}\hat{I}_-^2 +\text{h.c.},
\label{eq:HQ2}
\end{eqnarray}
where ${\mathbf P}$ is the nuclear quadrupole tensor and $P_{\alpha\beta}=\frac{Q}{2I(2I-1)}\langle\hat{V}_{\alpha\beta}\rangle$. Here $Q$ is the quadrupole constant that is 2507.2 (2648.2) mbarn for $^{161}$Dy ($^{163}$Dy) nucleus \cite{Pyykko2008}. $\langle\hat{V}_{\alpha\beta}\rangle$ is the expectation value of the electric-field gradient operator over the electronic ground doublet. $P_1=P_{xz}+iP_{yz}$, and $P_2=\frac{1}{2}\left(P_{xx}-P_{yy}\right)+iP_{xy}$.

\subsection{Computational Details}

Here we consider two different experimental geometries (Fig.~\ref{Geometry}): (i) DyPc$_2$ molecule {\it without} isotope enriched from Ref.~\cite{Marx2014} and (ii) $^{163}$DyPc$_2$ molecule with isotope enriched from Ref.~\cite{Pineda2017}. Henceforth, the former (latter) geometry is referred to as {\bf M1} ({\bf M2}). The structure of {\bf M1} is significantly deviated from $D_{4d}$ symmetry and it does not have any symmetry, whereas {\bf M2} has exact $C_4$ symmetry.

We perform the {\it ab-initio} calculations using the MOLCAS quantum chemistry code (version 8.2) \cite{Molcas} with the implementation of the hyperfine interactions as discussed in Ref.~\cite{Wysocki2019}. Scalar relativistic effects are considered in the form of Douglas-Kroll-Hess Hamiltonian \cite{Douglass1974,Hess1986}. For all atoms, relativistically contracted atomic natural orbital (ANO-RCC) basis sets are used \cite{Widmark1990,Roos2004}. For the Dy ion, we use polarized valence triple-$\zeta$ quality (ANO-RCC-VTZP), and for the N and C atoms, we use polarized valence double-$\zeta$ quality (ANO-RCC-VDZP). For the H atoms, we use valence double-$\zeta$ quality (ANO-RCC-VDZ). Our choice of the basis set is very similar to that used for DyPc$_2$ SMM in Ref.~\cite{Marx2014}.

In order to compute the electronic structure, we first apply state-averaged complete active space self-consistent (SA-CASSCF) method \cite{Roos1980,Siegbahn1981} to spin-free states, without SOI. We consider the complete active space consisting of only seven $f$ orbitals with nine electrons. Our previous calculation \cite{Pederson2019,Wysocki2019} on TbPc$_2$ SMMs showed that larger active space including ligand orbitals gives rise to a negligible effect on the low-energy electronic spectrum and hyperfine interaction parameters. With nine electrons on seven $f$ orbitals, there are 21 spin-free states (or roots) to build electron spin $S=5/2$. Once the state-average is performed over the 21 spin-free states, we include SOI within the atomic mean-field approximation \cite{Hess1996}, using the restricted active space state-interaction (RASSI) method \cite{rassi}. Then we extract the {\bf A} matrix from Eq.~(\ref{eq:Aformula}) and the {\bf P} matrix evaluated over the {\it ab-initio} electronic ground doublet.

\section{Results and Discussion}

\subsection{Electronic Energy Spectrum}

\begin{table}[t!]
\centering
\caption{Eigenvalues of the $\mathbf{g}$ matrix for the electronic ground doublet as well as the energy difference between the electronic ground and the first-excited doublet, $E_{\rm ZFS}$, for the anionic DyPc$_2$ SMMs for two different experimental geometries.}
\begin{tabular}{c|cccc}
\hline
\hline
 Geometry    & $g_{xx}$ & $g_{yy}$ & $g_{zz}$ & $E_{\rm ZFS}$~(cm$^{-1}$)\\
\hline
 {\bf M1} (Ref.~\cite{Marx2014})   & 0.0003 & 0.0003 & 17.4976 & 59.4 \\
 {\bf M2} (Ref.~\cite{Pineda2017}) & 0.0002 & 0.0002 & 17.3864 & 51.4 \\
\hline
\hline
\end{tabular}
\label{gMatrix}
\end{table}

Our {\it ab-initio} calculations shows that the ground multiplet $J=15/2$ is split into eight Kramers doublets due to the Pc ligands. For both experimental geometries, the electronic ground doublet $|g \rangle$ in the $J=15/2$ multiplet is well separated from the first-excited doublet $|e \rangle$ (Table~\ref{gMatrix}). For {\bf M1} [Fig.~\ref{Geometry}(a)], the doublet $|g \rangle$ consists of mainly $|M_J=\pm 13/2 \rangle$ with tiny contributions from $|M_J=\pm 15/2 \rangle$, and $|M_J=\pm 11/2 \rangle$, while the doublet $|e \rangle$ comprises mainly $|M_J=\pm 11/2 \rangle$ with very small contributions from $|M_J=\pm 15/2 \rangle$, and $|M_J=\pm 13/2 \rangle$. For {\bf M2} [Fig.~\ref{Geometry}(b)], the two doublets have pure $M_J$ states such as $|g \rangle=|M_J=\pm13/2\rangle$ and $|e \rangle=|M_J=\pm11/2\rangle$. The calculated $E_{\rm ZFS}$ value and the characteristics of the eigenstates agree well with the reported {\it ab-initio} results \cite{Marx2014}.

Since the energy gap between the ground and first excited Kramers doublets is much greater than the scale of the hyperfine interaction ($\sim0.1$ cm$^{-1}$), for the studies of the low-energy electronic-nuclear spectrum we can consider only the ground Kramers doublet (Fig.~\ref{Schematic}). The ground Kramers doublet can be represented by a fictitious pseudo-spin $S=1/2$ and the pseudo-spin formalism from the previous section can be used for description of the hyperfine coupling.

It is convenient to present the calculated $\mathbf{A}$ matrix and $\mathbf{P}$ tensor in the magnetic coordinate system in which the $\mathbf{g}$ matrix for the ground Kramers doublet is diagonal. The calculated eigenvalues of the $\mathbf{g}$ matrix are shown in Table~\ref{gMatrix} for both considered geometries. As expected, the $\mathbf{g}$ matrix is highly anisotropic with one large eigenvalue being approximately equal to $2g_J13/2\approx17.333$ (where $g_J\approx1.33$ is the Lande $g$ factor for Dy$^{+3}$ ion). The remaining two eigenvalues are very small but they are responsible for quantum tunneling of magnetization (QTM) process (which is discussed later). We choose the $z$ axis to point along the eigenvector corresponding to the large eigenvalue. This direction points approximately perpendicular to the ligand planes (see Fig.~\ref{Geometry}).

\subsection{Magnetic hyperfine interactions}

\begin{table}[t!]
\centering
\caption{Calculated Elements\textsuperscript{\emph{a}} of the Magnetic Hyperfine Matrix in Units of MHz for Two Dy Isotopes}
\begin{tabular}{l|cccccc|ccc}
\hline
\hline
 Isotope    & $A_{xx}$ & $A_{yy}$ & $A_{zz}$ & $A_{xy}$ & $A_{xz}$ & $A_{yz}$ & $|A_0|$    & $|A_1|$ & $|A_2|$ \\
\hline
 $^{161}$Dy {\bf M1} & -0.02 & -0.03 & -1444.29 & 0.00 &  1.16 & -0.60 & -0.02 & 0.65 & 0.00 \\
 $^{163}$Dy {\bf M1} &  0.03 &  0.04 &  2021.28 & 0.00 & -1.62 &  0.84 &  0.03 & 0.91 & 0.00 \\
 $^{163}$Dy {\bf M2} &  0.03 &  0.03 &  2005.30 & 0.00 &  0.00 &  0.00 &  0.03 & 0.00 & 0.00 \\
\hline
\hline
\end{tabular}
\\
\raggedright
\textsuperscript{\emph{a}} Used the magnetic coordinate system (Fig.~\ref{Geometry}) in which the ${\mathbf g}$ matrix for the electronic ground doublet is diagonal.
\label{AMatrix}
\end{table}

Table~\ref{AMatrix} shows the calculated elements of the magnetic hyperfine matrix for two $I=5/2$ Dy isotopes using both experimental geometries. For the \textbf{M2} geometry we only show the results for $^{163}$Dy since this isotope was solely used in the synthesis of \textbf{M2}. In both cases, the $A_{zz}$ element is dominant, while the other $\mathbf{A}$ matrix elements are close to zero. Similar behavior was found for the TbPc$_2$ molecule \cite{Wysocki2019}. However, an important difference is that for TbPc$_2$, the $A_{xx}$ and $A_{yy}$ elements are zero \cite{Wysocki2019}. This is a consequence of the fact that Tb$^{+3}$ is a non-Kramers ion and as a result only one eigenvalue of the $\mathbf{A}$ matrix is non-zero \cite{Griffith1963}. On the other hand, for DyPc$_2$ with Dy$^{+3}$ being a Kramers ion, all three $\mathbf{A}$ matrix eigenvalues are non-zero. This is reflected in non-zero $A_{xx}$ and $A_{yy}$ elements (see Table~\ref{AMatrix}). While $A_{xx}$ and $A_{yy}$ are very small ($<0.1$~MHz), they do have an important effect on an energy spectrum and magnetization dynamics (see below).

The presence of non-zero $A_{xz}$ and $A_{yz}$ elements for \textbf{M1} geometry is due to slight misalignment between the $z$ axes of the $\mathbf{g}$ matrix and $\mathbf{A}$ matrix coordinate systems. Such misalignment is possible when we have deviations from the $C_4$ symmetry and it originates from the interaction of the $J=15/2$ ground-multiplet with higher multiplets \cite{AbragamBook}. Since the \textbf{M2} geometry has the $C_4$ symmetry, the $\mathbf{g}$ matrix and $\mathbf{A}$ matrix coordinate systems are aligned and all off-diagonal elements of the $\mathbf{A}$ matrix are zero (see Table~\ref{AMatrix}).

Note that while the overall sign of the $A_{zz}$ element is undetermined in our calculations, the $^{161}$Dy and $^{163}$Dy isotopes have opposite sign of the $A_{zz}$ element. This is due to opposite sign of the nuclear $g$-factor for these isotopes. The difference in nuclear g-factors is also responsible for $^{163}$Dy having larger magnitude of $A_{zz}$.

\begin{figure}[h]
\centering
\includegraphics[width=0.7\linewidth]{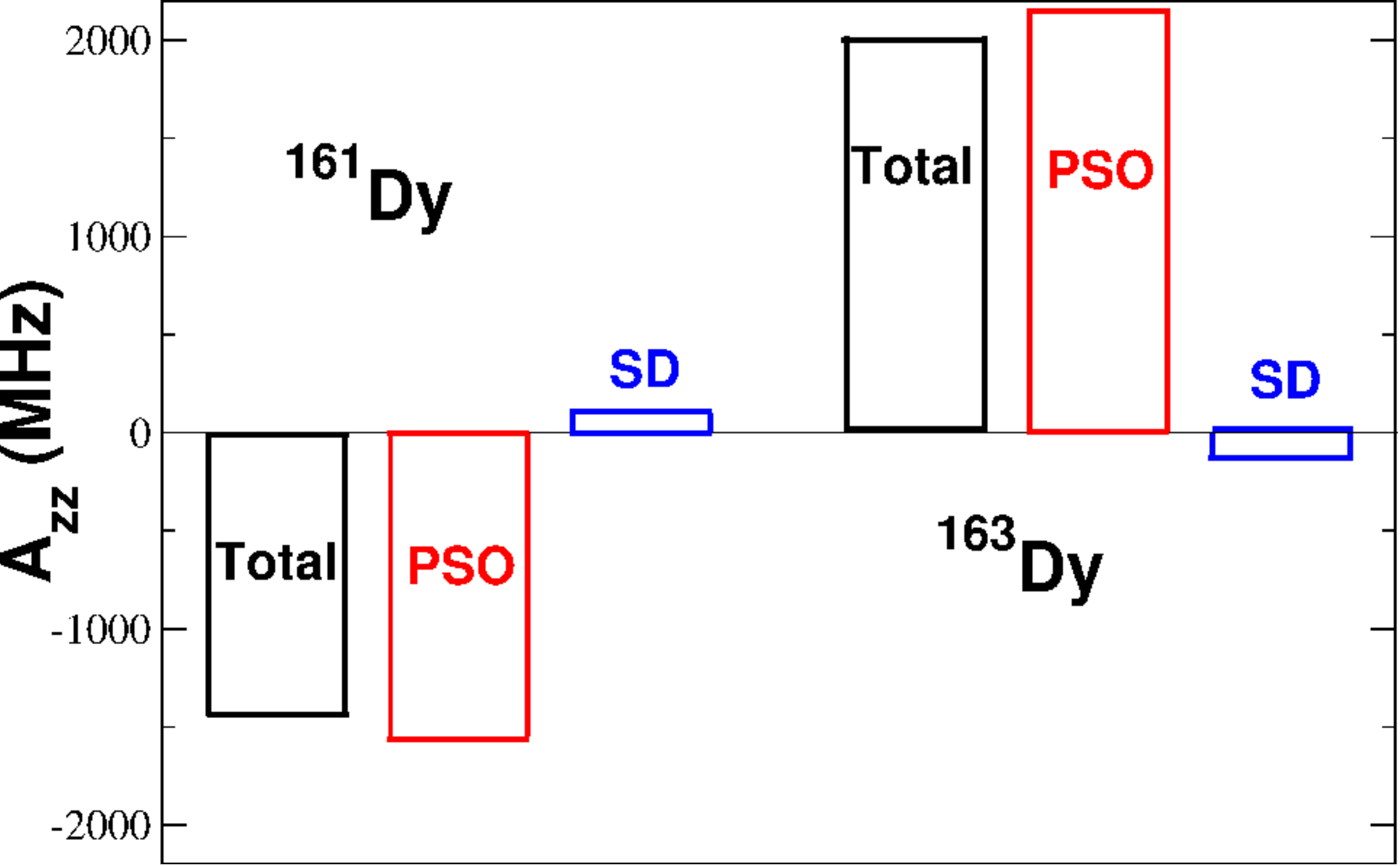}
\caption{Contributions of PSO and SD terms to the total calculated hyperfine parameter $A_{zz}$ for $^{161}$DyPc$_2$ and $^{163}$DyPc$_2$. Here the FC term is not shown since the magnitude is about 0.1~MHz for both Dy isotopes.}
\label{Histogram}
\end{figure}

Figure~\ref{Histogram} shows PSO and SD contributions to $A_{zz}$ compared with its total value. The FC contribution is negligible (less than 0.1~MHz) and is not shown. As in the case of the TbPc$_2$ molecule, the magnetic hyperfine interaction is dominated by the PSO mechanism due to Dy ion having a large orbital angular momentum. The much smaller SD contribution is opposite to the PSO part which results in the total $A_{zz}$ value being somewhat smaller than the PSO contribution.

To compare with the experimental value \cite{Pineda2017}, care needs to be exercised due to slightly different model Hamiltonians. The experimental quantity of $A_{\rm hf}^{\rm exp} J_z$ is equivalent to our calculated quantity of $A_{zz} S$, where $J_z=13/2$ and $S=1/2$ [electronic effective spin in Eq.~(\ref{eq:HA2})], ignoring the small non-axial hyperfine parameters. For $^{163}$DyPc$_2$ with \textbf{M2} geometry, the experimental value $A_{\rm hf}^{\rm exp}=153$~MHz with $J_z=13/2$ \cite{Pineda2017} is comparable to our calculated value of $A_{zz}\sim$2000~MHz with an effective electron spin $S=1/2$. Thus, we find good agreement between theory and experiment.

\subsection{Nuclear quadrupole interaction}

\begin{table}[t!]
\centering
\caption{Calculated Elements\textsuperscript{\emph{a}} of the Nuclear Quadrupole Tensor in Units of MHz for Two Dy Isotopes}
\begin{tabular}{l|cccccc|cc}
\hline
\hline
 Isotope & $P_{xx}$ & $P_{yy}$ & $P_{zz}$ & $P_{xy}$ & $P_{xz}$ & $P_{yz}$ & $|P_1|$ & $|P_2|$ \\
\hline
 $^{161}$Dy {\bf M1} & -47.7 & -47.2 & 95.0  & 0.9 & -3.9 & 1.3 & 4.1  & 0.9 \\
 $^{163}$Dy {\bf M1} & -50.4 & -49.9 & 100.3 & 0.9 & -4.1 & 1.4 & 4.4  & 1.0 \\
 $^{163}$Dy {\bf M2} & -50.6 & -50.6 & 101.2 & 0.0 &  0.0 & 0.0 & 0.0  & 0.0 \\
\hline
\hline
\end{tabular}
\\
\raggedright
\textsuperscript{\emph{a}} Used the magnetic coordinate system (Fig.~\ref{Geometry}) in which the ${\mathbf g}$ matrix for the electronic
ground doublet is diagonal.
\label{PMatrix}
\end{table}

The calculated elements of the nuclear quadrupole tensor are shown in Table~\ref{PMatrix} for the two considered isotopes using both experimental geometries. In all cases the uniaxial quadrupole parameter $P_{zz}$ is around 100~MHz. For \textbf{M1}, the transverse quadrupole parameters are of the order of few MHz. On the other hand, for \textbf{M2}, the transverse quadrupole parameters are identically zero due to $C_4$ symmetry.

To compare with the experimental value \cite{Pineda2017}, similarly to the case of the hyperfine interactions, we need conversion due to slightly different model Hamiltonians used in theory and experiment. The experimental parameter $P_{\rm exp}$ in Eq.~(1) in Ref.\cite{Pineda2017} is equivalent to $\frac{3}{2}P_{zz}$ in our formalism. The experimental value $P_{\rm exp}=420$~MHz obtained for $^{163}$DyPc$_2$ with \textbf{M2} geometry from fitting the observed steps in the magnetic hysteresis loops, is somewhat larger than our calculated value $\frac{3}{2}P_{zz}\sim$150~MHz. The effect of this discrepancy will be discussed below.

\begin{figure}[t!]
\centering
\includegraphics[width=0.8\linewidth]{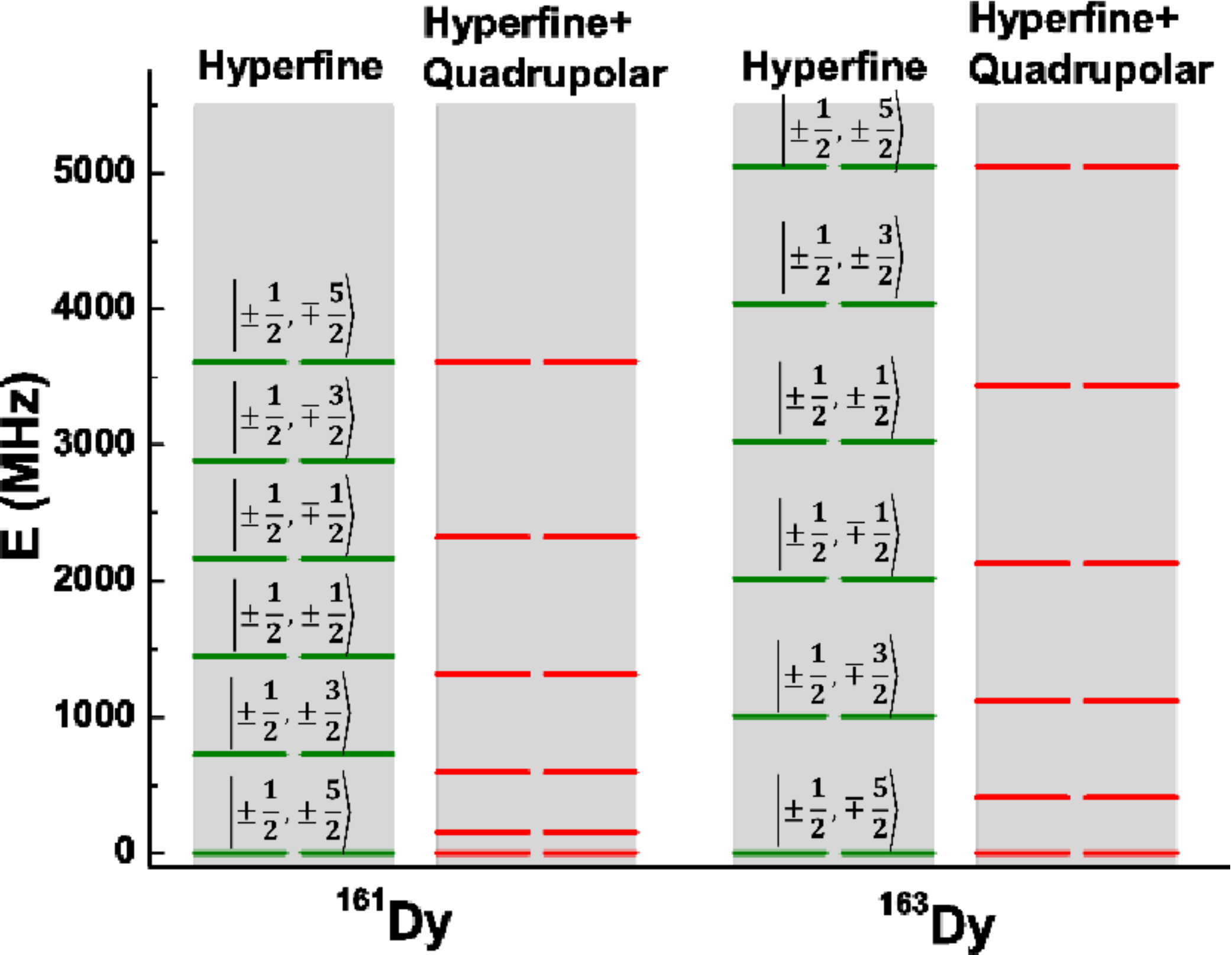}
\caption{The calculated low-energy electronic-nuclear spectra of the DyPc$_2$ molecule for two considered Dy isotopes with experimental \textbf{M1} geometry. The approximately equally spaced green lines correspond to energy levels found by diagonalization of the magnetic hyperfine Hamiltonian. The level characteristics are denoted. The red lines correspond to energy levels found by diagonalization of the sum of the magnetic hyperfine and nuclear quadrupole Hamiltonians. The $i^\text{th}$ red level have the same characteristic as the $i^\text{th}$ green level.}
\label{EnergyLevels}
\end{figure}

\subsection{Electronic-nuclear energy spectrum}

Using the calculated elements of the magnetic hyperfine matrix and the nuclear quadrupole tensor, we calculate the low-energy electronic-nuclear spectrum by diagonalizing the effective pseudo-spin Hamiltonian. The resulting energy levels are shown in Fig.~\ref{EnergyLevels} for both considered isotopes with the \textbf{M1} geometry. The spectrum is composed of six quasi-doublets.  Each doublet can be characterized by the $\ket{M_S,M_I}$ that has the largest contribution to the quasi-doublet (Fig.\ref{EnergyLevels}). Due to opposite signs of $A_{zz}$ for the two isotopes, the $^{161}$Dy and $^{163}$Dy nuclei have a reversed ordering of the doublet characters. In particular, for $^{161}$Dy, the ground doublet has a main contribution from the $\ket{\pm1/2,\pm5/2}$ states, while for $^{163}$Dy, the main contribution to the ground doublet comes from $\ket{\pm1/2,\mp5/2}$ states.

Unlike in the TbPc$_2$ case, the quasi-doublets have non-zero tunnel splittings due to presence of non-zero $A_{xx}$ and $A_{yy}$ elements. The largest tunnel splitting occurs for the $\ket{\pm1/2,\mp1/2}$ doublet ($\sim0.1$~MHz). In fact, for the \textbf{M2} geometry with the $C_4$ symmetry, the $\ket{\pm1/2,\mp1/2}$ doublet is the only doublet that has non-zero splitting. Deviations from the $C_4$ symmetry for the \textbf{M1} geometry, additionally, lead to splitting of the $\ket{\pm1/2,\pm1/2}$ doublet ($\sim0.01$~MHz), while tunnel splittings of other quasi-doublets being significantly smaller. Note that for $^{161}$Dy with $A_{zz}<0$, $\ket{\pm1/2,\mp1/2}$ is the 4$^\text{th}$ lowest quasi-doublet, whereas for $^{163}$Dy with $A_{zz}>0$, $\ket{\pm1/2,\mp1/2}$ is the 3$^\text{rd}$ lowest quasi-doublet (Fig.~\ref{EnergyLevels}). These tunnel splittings play an important role in magnetization dynamics (see below).

\begin{table}[t!]
\centering
\caption{Calculated and Experimental\textsuperscript{\emph{a}} Electronic-Nuclear Relative Energy Levels in GHz}
\begin{tabular}{l|ccc|c}
\hline
 Levels\textsuperscript{\emph{b}} & $^{161}$Dy ({\bf M1}) & $^{163}$Dy ({\bf M1}) & $^{163}$Dy ({\bf M2}) & Exp. $^{163}$Dy ({\bf M2}) \\
\hline
 $E_2-E_1$  & 0.156  & 0.410  & 0.395 & 0.5 \\
 $E_3-E_2$  & 0.436  & 0.709  & 0.699 & 0.7 \\
 $E_4-E_3$  & 0.722  & 1.010  & 1.003 & 1.0 \\
 $E_5-E_4$  & 1.007  & 1.312  & 1.306 & 1.3 \\
 $E_6-E_5$  & 1.292  & 1.613  & 1.610 & 1.5 \\
\hline
\end{tabular}
\\
\raggedright
\textsuperscript{\emph{a}} Extracted from Fig. 4c from Ref.~\cite{Pineda2017}. \textsuperscript{\emph{b}} $E_i$ denotes $i^{\text{th}}$ lowest electronic-nuclear doublet.
\label{Exp}
\end{table}

If the quadrupole interaction is neglected, the quasi-doublets are approximately equidistant with the energy gap of 722~MHz and 1011~MHz for $^{161}$Dy and $^{163}$Dy, respectively. The deviations from the equidistance are very small ($\sim0.01$~MHz) and are due to non-zero $A_{xx}$ and $A_{yy}$ elements. The larger gap for $^{163}$Dy is a consequence of larger $A_{zz}$ for this isotope. When the quadrupole coupling is included, the quasi-doublets are no longer equidistant and the gap between quasi-doublets increases for higher lying states.

Table~\ref{Exp} shows our calculated gaps between electronic-nuclear quasi-doublets for both isotopes considering both the hyperfine and quadrupole interactions. In Table~\ref{Exp} our calculations for $^{163}$Dy ({\bf M2}) are in a good agreement with experiment for $^{163}$Dy ({\bf M2}) from Ref.~\cite{Pineda2017}, considering typical experimental uncertainty such as about 0.1 GHz (see Ref.~\cite{Thiele2014}), as well as approximations made in our calculations (see Sec. 2.2). There are no reported experimental data for $^{161}$Dy ({\bf M1}) and $^{163}$Dy ({\bf M1}) molecules. Due to different geometries and different isotope species, we do not expect that the calculated values for $^{161}$Dy ({\bf M1}) and $^{163}$Dy ({\bf M1}) molecules are the same as that for $^{163}$Dy ({\bf M2}). 

%and compared with experimental values for $^{163}$Dy extracted from Ref.\cite{Pineda2017}. 
%Despite the  difference in the calculated and experimental quadrupole parameters, the calculated  level separations are in a good agreement wit%h the experimental values. The differences between experimental and calculated energy levels are due to (aforementioned) approximations made in% calculations as well as in experimental procedure.

\subsection{Zeeman diagram}

\begin{figure}[t!]
\centering
\includegraphics[width=0.8\linewidth]{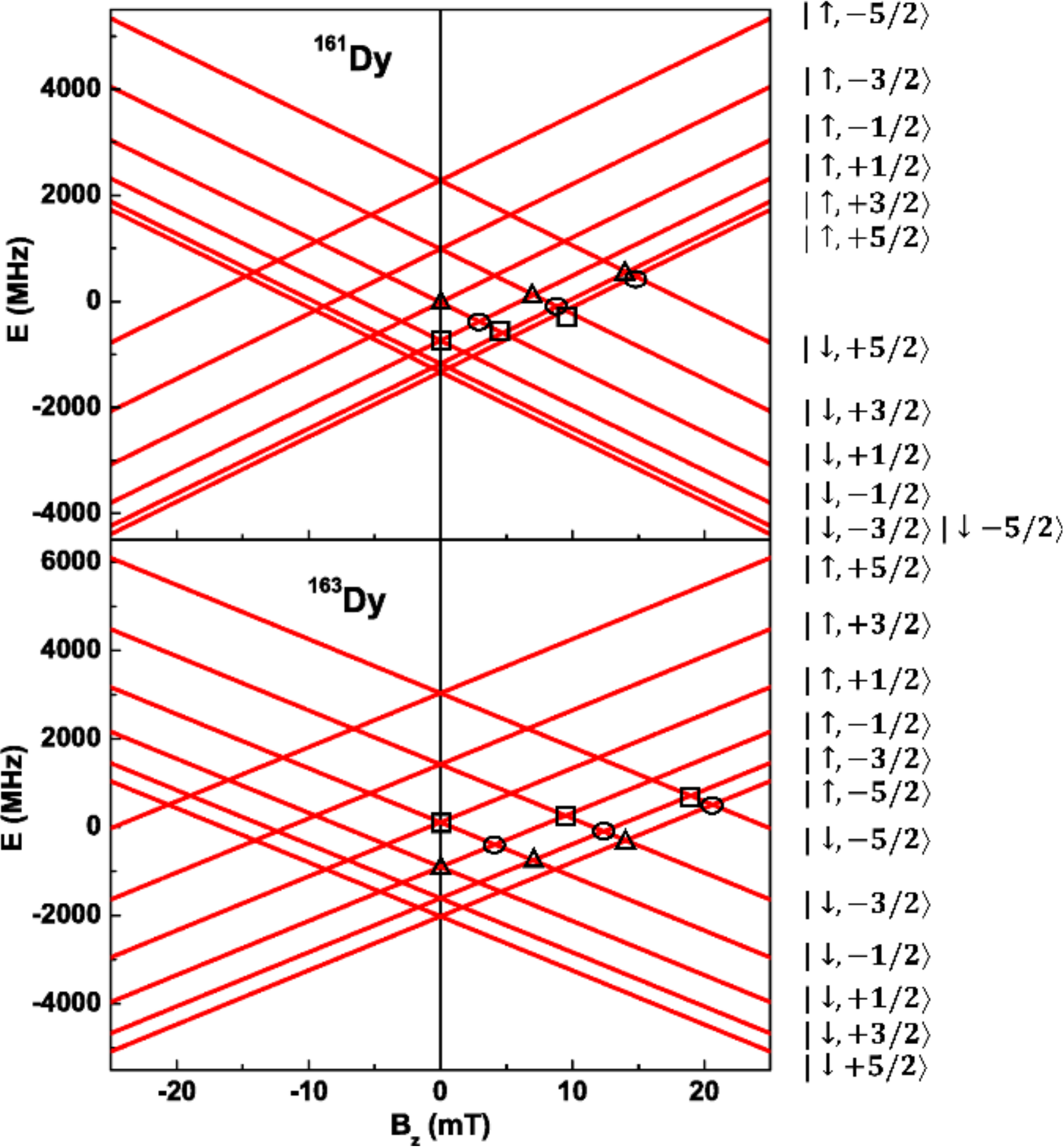}
\caption{Zeeman diagram showing the calculated electronic-nuclear energy levels as a function of magnetic field along the $z$ axis ($B_z$) for the DyPc$_2$ molecule with \textbf{M1} geometry for $^{161}$Dy (top) and $^{163}$Dy (bottom) isotopes. Here $\ket{M_S,M_I}$ represents the approximate quantum numbers of the levels. Open ovals, squares, and triangles denote avoided level crossing points with $\Delta M_I$ equal to 0, 1, and -1, respectively. See the main text for the definition of $\Delta M_I$. The crossing points for the negative magnetic field are not marked.}
\label{ZeemanDiagram}
\end{figure}

Let us now study how the electronic-nuclear energy levels vary in the presence of an external magnetic field along the $z$ axis ($B_z$). For this purpose we add the Zeeman pseudo-spin Hamiltonian $\hat{H}_{\text Z}=\mu_{\text B}B_{z}g_{zz}\hat{S}_z$ to the magnetic hyperfine and nuclear quadrupole terms [Eqs.~(\ref{eq:HA}) and~(\ref{eq:HQ})] and diagonalize the resulting Hamiltonian as a function of $B_z$. The resulting Zeeman diagram is shown in Fig.~\ref{ZeemanDiagram} for two considered Dy isotopes using the \textbf{M1} geometry. The Zeeman diagrams for $^{163}$Dy isotope for the \textbf{M1} and \textbf{M2} geometries are similar to each other. Note that the diagram is symmetric with respect to $B_z\rightarrow -B_z$. The energy levels can be denoted by the approximate $M_S$ and $M_I$ quantum numbers (see the right hand side of Fig.~\ref{ZeemanDiagram}). All the levels with $M_S=\uparrow$ linearly vary with magnetic field with the same positive slope, while the levels with $M_S=\downarrow$ vary with the opposite slope.

At certain magnetic field values, the levels with opposite $M_S$ appear to cross each other. A crossing point can be characterized by the difference between $M_I$ of the two crossing levels, $\Delta M_I=M^\uparrow_I-M^\downarrow_I$. Here $M^\uparrow_I$ ($M^\downarrow_I$) is the $M_I$ value for the electronic-nuclear level $|M_S=\uparrow(\downarrow),M_I\rangle$. Some of these apparent crossing points can be split and become avoided level crossings (ALCs). The ALCs play an important role in magnetization dynamics since in their proximity QTM processes are possible. (QTM can be used to read the state of the nuclear spin levels for quantum information applications \cite{Thiele2014}.) 
For TbPc$_2$ molecules, ALCs (or steps in magnetic hysteresis) exist only at non-zero magnetic fields and are caused by transverse CF interactions \cite{Wysocki2019,Ishikawa2005}. In this case, ALCs with $\Delta M_I=0$ are mainly responsible for QTM (with much smaller contributions from $\Delta M_I=\pm1,\pm2$ \cite{Wysocki2019,Taran2019}). For DyPc$_2$ molecules, however, situation is quite different. Here, due to non-zero $A_{xx}$ and $A_{yy}$ elements, magnetic hyperfine interactions, in general, give rise to tunnel splitting at crossing points with $\Delta M_{\text{ALC}}=\pm1$ (squares and triangles in Fig.~\ref{ZeemanDiagram}) with and without $B_z$ field. Importantly, tunnel splitting at crossing points with $\Delta M_{\text{ALC}}=-1$ (triangles in Fig.~\ref{ZeemanDiagram}) remains non-zero even for the $C_4$ symmetry. On the other hand, non-zero splitting at crossing points with $\Delta M_{\text{ALC}}=1$ (squares in Fig.~\ref{ZeemanDiagram}) requires deviations from the $C_4$ symmetry. Therefore, for DyPc$_2$, QTM is possible even at zero magnetic field, which is in agreement with experiment \cite{Pineda2017,Ishikawa2005}.

Crossing points with $\Delta M_I=0$ (ovals in Fig.~\ref{ZeemanDiagram}) can become ALCs in the presence of additional small transverse magnetic field (not included in calculations). Such field can originate from hyperfine interactions with C and N nuclei or from dipolar interactions with different magnetic molecules. In the presence of non-zero transverse quadrupole parameters, transverse magnetic field can also induce tunnel splitting at crossing points with $\Delta M_I=\pm2$ (not shown) and further increase tunnel splitting at ALCs with $\Delta M_{\text{ALC}}=\pm1$. The splittings induced by the transverse quadrupole interactions are, however, very small.

\begin{table}[t!]
\centering
\caption{Magnetic field values (in mT) of ALCs for the \textbf{M1} geometry for two considered Dy isotopes.}
\begin{tabular}{c|cc}
\hline
 $\Delta M_I$ & $B_{\text{ALC}}$ $^{161}$Dy &  $B_{\text{ALC}}$ $^{163}$Dy \\
\hline
  0 & 14.8 & 20.6 \\
  0 &  8.8 & 12.4 \\
  0 &  3.0 &  4.1 \\
\hline
 +1 &  9.5 & 19.0 \\
 +1 &  4.7 &  9.5 \\
 +1 &  0.0 &  0.0 \\ 
 \hline
 -1 & 14.1 & 14.1 \\
 -1 &  7.1 &  7.0 \\
 -1 &  0.0  & 0.0 \\ 
\hline
\end{tabular}
\label{ALC}
\end{table}

Table~\ref{ALC} shows non-negative magnetic field values of ALCs with $\Delta M_I=0,\pm1$ for DyPc$_2$ molecule with geometry \textbf{M1} for $^{161}$Dy and $^{163}$Dy isotopes. A hysteresis loop of the DyPc$_2$ molecule is expected to show steps at the field values listed in Table~\ref{ALC}. For diluted crystals with smaller dipolar interactions, the largest steps are expected to occur at field values corresponding to ALCs with $\Delta M_I=-1$. However, the step size at ALCs with $\Delta M_I=0$ can be potentially tuned by application of a small transverse magnetic field.

\section{Conclusion}

Magnetic hyperfine and nuclear quadrupole interactions for anionic $^{161}$DyPc$_2$ and $^{163}$DyPc$_2$ SMMs with asymmetric and $C_4$ symmetric experimental geometries are investigated using multiconfigurational {\it ab-initio} methods combined with an effective Hamiltonian for an electronic ground Kramers doublet. For both geometries and both Dy isotopes, our calculations reveal that the hyperfine and quadrupole interactions are much smaller than those for $^{159}$TbPc$_2$ SMMs. In the case of the DyPc$_2$ SMMs, the hyperfine interactions can induce tunnel splitting at avoided level crossings even in the absence of an external magnetic field, which corroborates the presence of steps at zero magnetic field in observed magnetic hysteresis loops \cite{Pineda2017,Ishikawa2005}. This is due to the fact \cite{Griffith1963} that the hyperfine interactions for electronic Kramers doublets can have non-zero transverse parameters like $A_{xx}$ and $A_{yy}$, in contrast to the case of electronic non-Kramers quasi-doublets.

\section*{Acknowledgments}
This work was funded by the Department of Energy (DOE) Basic Energy Sciences (BES) grant No DE-SC0018326. Computational support by Virginia Tech ARC and San Diego Supercomputer Center (SDSC) under DMR060009N.

\section*{References}

\bibliography{refs}

\end{document}